\documentclass[epjST]{svjour}
\usepackage{graphics}
\usepackage{amsmath,amssymb}
\begin{document}
\title{Bifurcation analysis and chaos control of periodically driven discrete fractional order memristive Duffing Oscillator}
\author{Samuel Ogunjo\inst{1}\fnmsep\thanks{Corresponding author: \email{stogunjo@futa.edu.ng}} \and Ibiyinka Fuwape\inst{1,2}\thanks{\email{iafuwape@futa.edu.ng; iafuwape@mciu.edu.ng}} }
\institute{Department of Physics, Federal University of Technology, Akure, Ondo State, Nigeria. \and Michael and Cecilia Ibru University, Ughelli, Delta State, Nigeria.}
\abstract{
Discrete fractional order chaotic systems extends the memory capability to capture the discrete nature of physical systems.  In this research, the memristive discrete fractional order chaotic system is introduced.  The dynamics of the system was studied using bifurcation diagrams and phase space construction.  The system was found chaotic with fractional order $0.465<n<0.562$.   The dynamics of the system under different values makes it useful as a switch. Controllers were developed for the tracking control of the two systems to different trajectories.  The effectiveness of the designed controllers were confirmed using simulations.
} 
\maketitle
\section{Introduction}\label{intro}



The study of chaos arose from the study of integer order systems \cite{lorenz1963}.   This was soon extended to discrete systems to explain the behaviour in natural systems and models especially population models \cite{fuwapebifurcation}.  Many methods used in integer order and discrete systems have been developed over the years to study observational data such as climate \cite{Fuwape2017},  economic and financial data \cite{Lahmiri2017} and communication systems \cite{Fuwape2016}.  Although it has the capability to capture memory effect in models, fractional order systems have not gained attention over the years due to inadequate solution technique \cite{fuwapebifurcation}.   New developments in recent times such as computation methods and computing powers have brought attention to fractional order systems \cite{Shukla2017}.

There exist situations where it is desired that systems be chaotic.  This can be achieved through chaotification or anti-control \cite{qi}.  At other times when the existence of chaos in a system is undesirable, the methods of chaos control can be used to remove the effect of chaos in such system.   Chaos control involves obtaining a chaotic, periodic or stationary behaviour in a chaotic system by the application of tiny perturbations or controllers \cite{fuwapebifurcation,Boccaletti2000}.  Synchronization is a form of control whereby the trajectory of a chaotic system is made to track the trajectory of another system.  Different forms of synchronization such as projective synchronization, function projective synchronization, phase synchronization, lag synchronization, complete synchronization, anticipated synchronization have been proposed \cite{ojo2013comparison}.  Synchronization of chaotic systems has been found useful in secure communication \cite{ojo2012synchronization,Essimbi2012}.

The practical application of chaos to secure communication involves building electronic circuits for its implementation.  Chua \cite{chua1971} proposed a nonlinear device, the memristor, as the fourth circuit element in addition to the resistor, capacitor and inductor.  The memristor is considered as a nonlinear resistor with memory \cite{Danca2016}.  Novel chaotic systems which have been developed based on the memristor include the two component circuit \cite{Tchitnga2012}, memristor biomembranes \cite{Volkov2016}, memristive oscillatory systems \cite{Talukdar2012}, and Muthuswamy and Chua system \cite{muthuswamy}.  Memristors have practical potential applications in high performance computing, dynamic memory elements and neural synapses \cite{Sabarathinam}.

Discrete fractional order implementation of systems have been investigated including: Extension of integer order chaotic systems have been carried for systems such as  sine and standard maps \cite{wu2014discrete}, Logistic map \cite{wu2014chaos}, Chua system \cite{Agarwal2013} and Henon map \cite{hu2014discrete,liu2016chaotic}.  In this paper, we extend the study of the Duffing oscillators to the discrete fractional order model.  By introducing a memristor into the system, the different responses of the system to varying parameters were investigated using bifurcation diagrams.  Control of the memristive discrete fractional order model was also carried out using the method of active control.
\section{Methods}\label{methods}

\subsection{Memristors}\label{mem}
Different forms of flux dependent rate of change of charge have been proposed.  In this work, we study a flux controlled memristor of the form
\begin{equation}\label{flux}
    \phi(q) = \omega_0^2q + \beta q^3
\end{equation}
The memductance is obtained as
\begin{equation}\label{mem_def}
    M(q) = \frac{d\phi(q)}{dq} = \omega_0^2 + 3\beta q^2
\end{equation}
where $\omega_0^2$ and $\beta$ are constants.

\subsection{Fractional order systems}\label{frac_system}
The fractional sum of order $\nu < 0$ can be defined as \cite{atici2009}
\begin{equation}\label{atici}
    \Delta_a^{-\nu} x(t) := \frac{1}{\Gamma(\nu)}\sum_{s=a}^{t-\nu} (t-\sigma(s))^{(\nu-1)x(s)},\,\, t\in \{a,a+1,a+2,\ldots\}
\end{equation}
$\sigma(s) = s + 1$, and $t^{(\nu)}$ is the falling function defined as
\begin{equation}\label{falling_function}
    t^{(\nu)} = \frac{\Gamma(t+1)}{\Gamma(t+1-\nu)}
\end{equation}

This definition was extended by \cite{Abdeljawad2011} for the $\nu$-order ($\nu >0, \nu \not\in \mathbb{N}$) Caputo-like delta difference as
\begin{equation}\label{abdeljawad}
    \Delta_C^\nu x(t) := \frac{1}{\Gamma(n-\nu)}\sum_{s=a}^{t-(n-\nu)} (t-\sigma(s))^{(n-\nu-1)}\Delta^n_s x(s)
\end{equation}
where $n = [\nu] +1$.

Using this transformation, a discrete fractional system can be written as \cite{fuwapebifurcation}

\begin{equation}\label{caputo}
    \Delta^\nu_a x(t) = f(t+v-1,x(t+\nu -1)
\end{equation}
This has a solution given by
\begin{equation}\label{caputo2}
    x(t) = x_0 + \frac{1}{\Gamma(a)}\sum_{s=1-\nu}^{t-\nu} (t-s-1)^{(\nu -1)}f(s+\nu -1,x(s+v-1)
\end{equation}
\subsection{System description}\label{sys}
Duffing oscillator is defined as

\begin{equation}\label{duffing_integer}
    m\ddot{x} + c\dot{x} + kx + ax^3 = f\sin(\varpi t)
\end{equation}
where $m$, $c$, $k$, and $a$ are system constants.  Substituting Equation \ref{mem_def} and rewriting as a two dimensional system, the integer order Duffing oscillator can be written as
\begin{equation}\label{duffing_1}
    \begin{split}
      \dot{x} &= y \\
      \dot{y} &= f_1\sin(\varpi t) - M(q)x - f_2 y
    \end{split}
\end{equation}
where $f_1 = \frac{f}{m}$, $3\beta  = \frac{a}{m}$, $\omega_0^2 = \frac{k}{m}$, $M(q) = \omega_0^2 + 3\beta x^2$ and $f_2 = \frac{c}{m}$.  System \ref{duffing_1} can be written in fractional order form as
\begin{equation}\label{duffing_2}
    \begin{split}
      \frac{d^n x}{dt^n} &= y \\
      \frac{d^n y}{dt^n} &= f_1\sin(\varpi t) - M(q)x - f_2 y
    \end{split}
\end{equation}
where $n$ is the order of the system.   Using the transformation described in section \ref{frac_system}, the discrete fractional form of the memristive Duffing oscillator is written as
\begin{equation}\label{discrete_fractional}
    \begin{split}
      \Delta^\nu_a x & = y(t+\nu -1) \\
      \Delta^\nu_a y & = f_1\sin(\varpi t) - M(q)x(t+\nu -1) - f_2 y(t+\nu -1)
    \end{split}
\end{equation}
Sabarathinam et al \cite{Sabarathinam} transformed equation \ref{duffing_1} into a four dimensional integer order system for analysis.
\section{Results and discussion}

\subsection{Bifurcation analysis}
\begin{figure}
\centering
\resizebox{0.85\columnwidth}{!}{%
\includegraphics{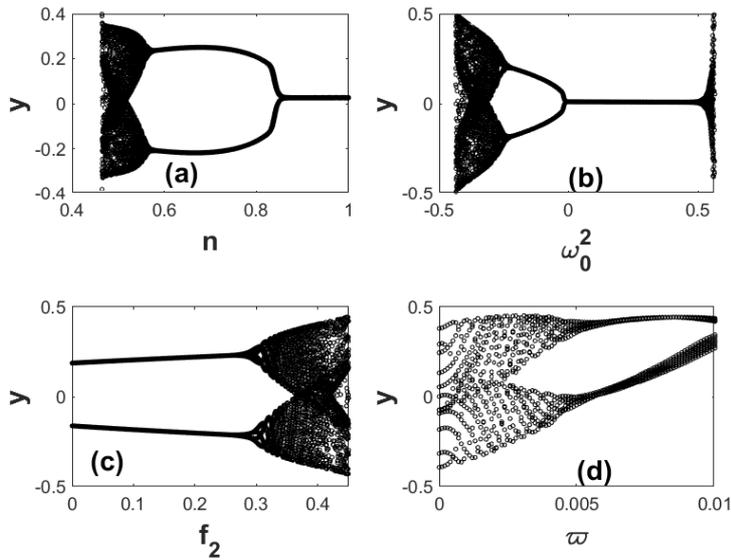} }
\caption{Bifurcation diagram for y-component with (a) fractional order, $n$ with parameters $(f_1,\omega_0^2,3\beta,f_2,\varpi) = (0.9, -0.35, 1.5, 0.4, 0.0002)$. (b) $\omega_0^2$ with system parameters  $(f_1,n,3\beta,f_2,\varpi)= (0.5, 0.5, 1, 0.4, 0.0002)$  (c) parameter, $f_2$ using $(f_1,\omega_0^2,3\beta,n,\varpi) = (0.5, -0.35, 1, 0.5, 0.0002)$; and (d) $\varpi$ with system parameters $(f_1,\omega_0^2,3\beta,f_2,n)=(0.5, -0.35, 1, 0.4, 0.5)$}
\label{fig:1}       
\end{figure}

\begin{figure}
\centering
\resizebox{0.75\columnwidth}{!}{%
\includegraphics{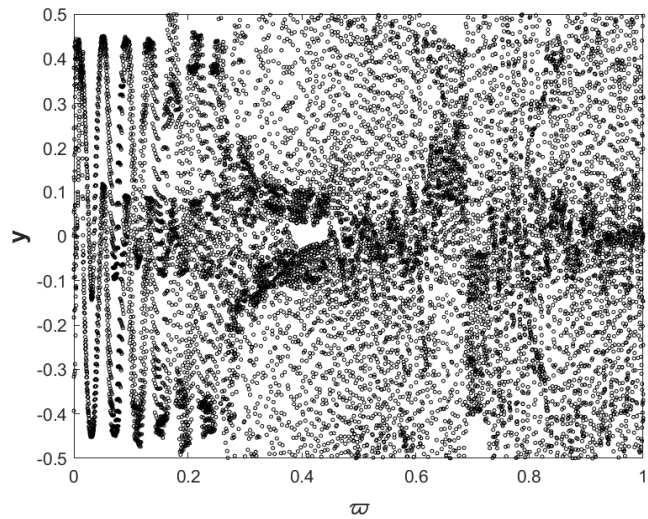} }
\caption{Bifurcation diagram for y-component with $\varpi$ in the range $0<\varpi<1$}
\label{fig:2}       
\end{figure}

\begin{figure}
\centering
\resizebox{0.75\columnwidth}{!}{%
\includegraphics{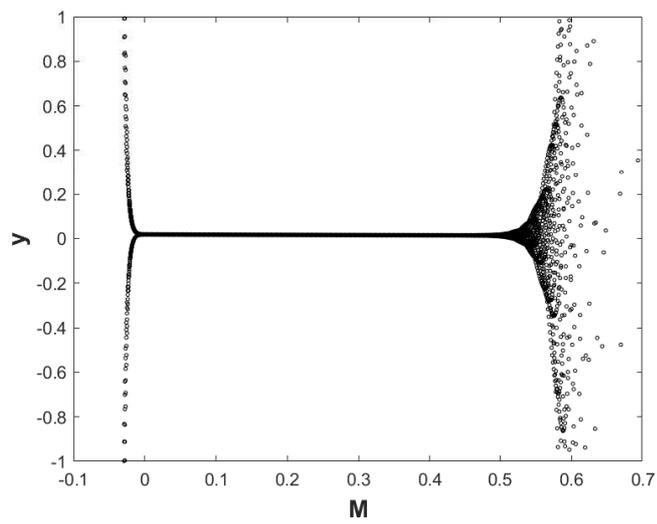} }
\caption{Bifurcation diagram for y-component with the memristor $M(q)$}
\label{fig:3}       
\end{figure}

\begin{figure}
\centering
\resizebox{0.75\columnwidth}{!}{%
\includegraphics{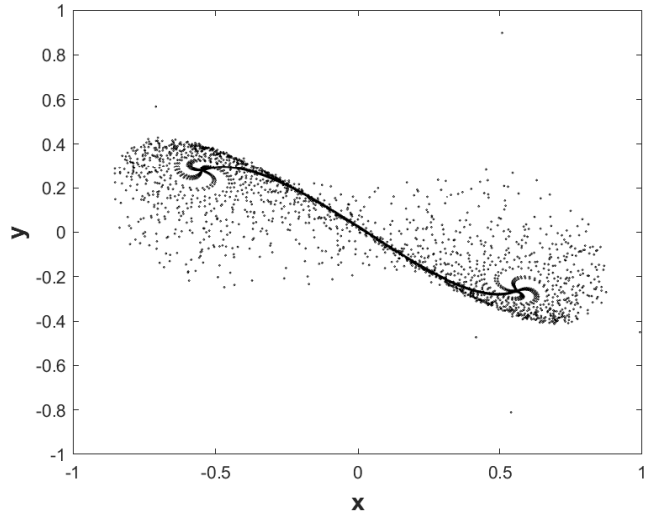} }
\caption{$x-y$ phase space realization of the driven Duffing oscillator obtained with parameter $(f_1,\omega_0^2,3\beta,f_2,\varpi) = (0.9, -0.35, 1.5, 0.4, 0.0002)$ and $0<n<1$.}
\label{fig:4}       
\end{figure}

The bifurcation analysis of the discrete fractional order Duffing oscillator with respect to different system parameters was investigated and the results presented in Figure \ref{fig:1}.  The system behaviour with respect to the fractional order, $n$, shows a reverse period doubling route to chaos (Figure \ref{fig:1}a).   Chaotic regions were found in the region $0.465<n<0.562$ with transition to period-2 and period-3 bifurcation at $n=0.86$ and $n=0.57$ respectively (Figure \ref{fig:1}b).   A period doubling route to chaos was also observed when the parameter $\omega_0^2$ was varied.  The system was found to be chaotic in only small positive region $0.517<\omega_0^2<0.566$.  Using system parameters $(f_1,n,3\beta,f_2,\varpi)= (0.5, 0.5, 1, 0.4, 0.0002)$, the Duffing oscillator with periodic forcing was found to be chaotic when $-0.43<\omega_0^2<-0.232$.  By varying the parameter $f_2$, the system was found to be chaotic when $0.288<f_2<0.45$ (Figure \ref{fig:1}c).  The parameter $\varpi$ presents an interesting bifurcation structure in the region $0<\varpi<1$.  In the small region $0<\varpi<0.01$, the system was found to be chaotic with a reverse period doubling bifurcation.  In Figure \ref{fig:2}, the system showed periodic behaviour in the region $0.01<\varpi<0.2$ beyond which it showed unusual chaotic behaviour.  Thus, we can conclude that the system follows a period doubling route to chaos when $0<\varpi<0.01$ but a quasiperiodic route when $\varpi>0.01$).  The bifurcation diagram obtained by varying the memristor in the system is shown in Figure \ref{fig:3} and the phase space when the fractional order is varied is presented in Figure \ref{fig:4}.  The behaviour of the memristor at different values suggest possible applications as a switch.

\subsection{Chaos control}
According to Fuwape and Ogunjo \cite{fuwapebifurcation}, tracking control is obtained when the individual components of a system followed different predefined rules.  The aim is to control the $x$ and $y$ component of System (\ref{discrete_fractional}) to follow different trajectories.  Rewriting the memristive discrete fractional order system (System \ref{discrete_fractional}) and adding controllers, we have
\begin{equation}\label{control0}
    \begin{split}
      \Delta^\nu_a x & = y(t+\nu -1) + u_1 \\
      \Delta^\nu_a y & = f_1\sin(\varpi t) - M(x)x(t+\nu -1) - f_2 y(t+\nu -1) + u_2
    \end{split}
\end{equation}
The error term is written as
\begin{equation}\label{control1}
    \begin{split}
      \dot{e}_1 &= \dot{x} - \dot{c}_1 \\
      \dot{e}_2 &= \dot{y} - \dot{c}_2
    \end{split}
\end{equation}
where $c(c_1,c_2)$ are user defined functions.   If $c_1\neq c_2$, mixed tracking is obtained.  Substituting Equation \ref{control1} into Equation \ref{discrete_fractional}, we obtain
\begin{equation}\label{control2x}
    \begin{split}
      \dot{e}_1 &= e_2 + c_2 - c_1 \\
      \dot{e}_2 &= f_1\sin(\varpi t) - \omega_0^2 e_1 - \omega_0^2 c_1 - 3\beta x^3 - f_2e_2 - f_2c_2 - c_2
    \end{split}
\end{equation}

Eliminating nonlinear terms in $e_1,\,e_2$ gives the subcontroller functions
\begin{equation}\label{control2}
    \begin{split}
      u_1 &= -c_2 + c_1 + v_1 \\
      u_2 &= -f_1\sin(\varpi t) + \omega_0^2 c_1 + 3\beta x^3 + f_2 c_2 + c_2 + v_2
    \end{split}
\end{equation}
where the parameters $v_i$ will be obtained later.  Substituting Equation \ref{control2} into Equation \ref{control2x}, we obtain

\begin{equation}\label{control3}
    \begin{pmatrix}
      \dot{e}_1 \\
      \dot{e}_2 \\
    \end{pmatrix}
    =
    \begin{pmatrix}
      0 & -1 \\
      \omega_0^2 &   f_2 \\
    \end{pmatrix}
     \begin{pmatrix}
      e_1 \\
      e_2 \\
    \end{pmatrix}
    +
     \begin{pmatrix}
      v_1 \\
      v_2 \\
    \end{pmatrix}
\end{equation}
The method of active control requires that a constant matrix $\mathbb{K}$ is chosen which will control the error dynamics such that the feedback matrix becomes
\begin{equation*}
    \begin{pmatrix}
      v_1 \\
      v_2 \\
    \end{pmatrix}
    = \mathbb{K} \begin{pmatrix}
      e_1 \\
      e_2 \\
    \end{pmatrix}
\end{equation*}
Thus, a matrix of the form
\begin{equation}\label{control4}
   \mathbb{K} =  \begin{pmatrix}
      \lambda_1 & -1 \\
      \omega_0^2 & \lambda_2 + f_2 \\
    \end{pmatrix}
\end{equation}

\begin{figure}
\centering
\resizebox{0.75\columnwidth}{!}{%
\includegraphics{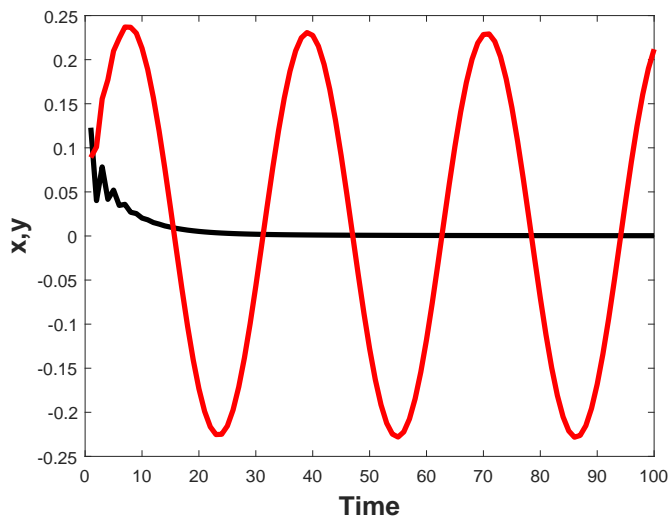} }
\caption{Tracking control of the driven Duffing oscillator obtained with parameter $(f_1,\omega_0^2,3\beta,f_2,\varpi,n) = (0.9, -0.35, 1.5, 0.4, 0.0002,0.5)$. The $x$ component (red line) is made to track the function $A\sin(wt)$ and the $y$ component (black line) tracks the function $Ae^{kt}$, where $A,w,k$ are constants.}
\label{fig:5}       
\end{figure}

To verify the effectiveness of the tracking controllers designed above, numerical simulations were carried out.  The $x$ component was made to track and exponential function while the $y$ component tracked a sine wave.  The result obtained is shown in Figure \ref{fig:5}.  The trajectories of the $x$ and $y$ component were found to track the desired set functions, hence, we conclude that tracking control of the system has been achieved.

\section{Conclusion}
Recently, there has been renewed interest in fractional order systems.  In this research, we extend the chaotic fractional order Duffing system to a discrete-fractional order system with a memristor.  Bifurcation diagrams were produced to show the different behaviour of the system under different parameter changes.  The possibility of control the two state space of the discrete fractional order memristive Duffing oscillator to two different trajectory was investigated.  The effectiveness of the controllers were demonstrated by numerical simulations.

\end{document}